\documentclass[aps,prl,twocolumn,noshowpacs,noshowkeys,superscriptaddress
]{revtex4-1}
\usepackage{graphicx}
\usepackage{textcomp}
\usepackage{amsfonts}
\usepackage{amsmath}
\usepackage{amssymb}
\usepackage{multirow}
\usepackage{epstopdf}

\begin{document}

\title{A single NV defect coupled to a nanomechanical oscillator}

\author{O. ~Arcizet}
\email{olivier.arcizet@grenoble.cnrs.fr}
\homepage[]{www.neel.cnrs.fr}
\affiliation{Institut N\'{e}el, CNRS et Universit\'{e} Joseph Fourier, 38042 Grenoble Cedex 09, France}
\author{V. ~Jacques}
\affiliation{Laboratoire de Photonique Quantique et Mol\'{e}culaire, 94235 Cachan, France}
\author{A. ~Siria}
\author{P. ~Poncharal}
\author{P. ~Vincent}
\affiliation{Laboratoire de Physique de la Mati\`{e}re Condens\'ee et Nanostructures, CNRS and Universit\'{e} Claude Bernard, 69622 Villeurbanne, France}
\author{S. ~Seidelin}
\affiliation{Institut N\'{e}el, CNRS et Universit\'{e} Joseph Fourier, 38042 Grenoble Cedex 09, France}

\begin{abstract} {A single Nitrogen Vacancy (NV) center hosted in a diamond nanocrystal is positioned at the extremity of a SiC nanowire. This novel hybrid system couples the degrees of freedom of two radically different systems, i.e. a nanomechanical oscillator and a single quantum object. The dynamics of the nano-resonator is probed through time resolved nanocrystal fluorescence and photon correlation measurements, conveying the influence of a mechanical degree of freedom given to a non-classical photon emitter.
Moreover, by immersing the system in a strong magnetic field gradient, we induce a magnetic coupling between
the nanomechanical oscillator and the NV electronic spin, providing nanomotion readout through a single electronic spin.
Spin-dependent forces inherent to this coupling scheme are essential in a variety of active cooling and entanglement protocols
used in atomic physics, and should now be within the reach of nanomechanical hybrid systems.}\end{abstract}
\maketitle

Owing to recent developments in cavity opto- and electro-mechanics \cite{Aspelmeyer2008, Kippenberg2008, Schwab2005}, it is now realistic to envision the observation of macroscopic mechanical oscillators cooled by active or traditional cryogenic techniques close to their ground state of motion. This conceptually elegant accomplishment would give access to a vast playground for physicists if
the resonator wavefunction could be coherently manipulated such as to create, maintain and probe Fock or other non-classical states. It would provide a remarkable opportunity to extend the pioneering experiments with trapped ions \cite{Blatt2008} to encompass macroscopic objects.
However, standard continuous measurements techniques used to actively cool and probe the resonator \cite{Braginsky1992}, when utilized to manipulate its quantum state, tend to blur its non-classical nature. An attractive alternative consists in interfacing the mechanical degrees of freedom with a single quantum object such as a 2-level system whose quantum state can be externally controlled \cite{Wilson-Rae2004,Hammerer2009, Rabl2009,Hunger2010,LaHaye2009,Bennett2010}.  Successful realization of this type of coupling  between a nanomechanical oscillator in the quantum regime and a phase qubit was recently reported \cite{O'Connell2010} and motivates the development of similar hybrid quantum systems presenting extended coherence times at room temperature and compatible with continuous measurement approaches.

\begin{figure}[b]
\begin{center}
\includegraphics[width=8.3cm]{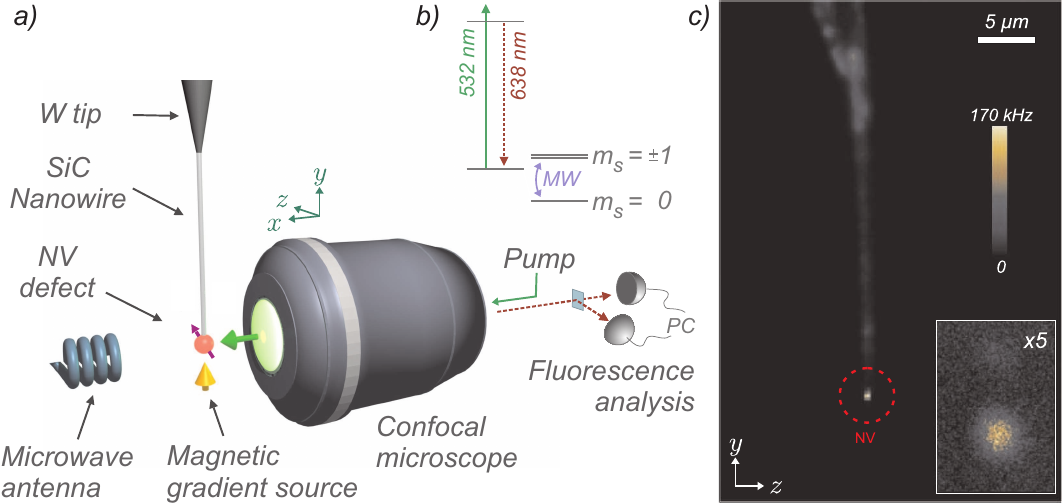}
\caption{\small  The hybrid system. (a): A confocal microscope monitors the fluorescence of a single NV defect hosted in a diamond nanocrystal positioned at the extremity of a SiC nanowire. A microwave antenna is used to manipulate the NV electronic spin, while a micro-fabricated magnetic structure approached in the vicinity of the suspended NV center generates a strong magnetic field gradient. (b): Simplified electronic structure of the NV centers at zero magnetic field. (c): Fluorescence map of the system recorded with the confocal microscope while scanning the objective position. The isolated bright spot circled in red corresponds to the fluorescence of a single NV center.  Inset: zoom on the nanowire extremity.
}
\end{center}
\end{figure}

Here we report a first step in this direction by coupling a nanomechanical oscillator to a single negatively-charged Nitrogen Vacancy (NV) defect hosted in a diamond nanocrystal attached to its extremity (fig 1a). In that context, the NV defect appears as an attractive quantum system, both for its optical and electronic spin properties. Indeed,  perfect photostability at room temperature makes the NV defect a robust and practical single-photon source \cite{Kurtsiefer2000,Brouri2000}. Moreover, the NV defect ground state is a spin triplet (fig 1b) which can be initialized and read-out by optical means, and manipulated by resonant microwave excitation with an unprecedented coherence time for a solid-state system under ambient conditions \cite{Jelezko2004,Balasubramanian2009}. Such properties are at the heart of diamond-based quantum information processing \cite{Gurudev2007,Neumann2008,Buckley2010,Togan2010,Neumann2010} and ultrasensitive magnetometry, where the spin is used as an atomic sized magnetic sensor \cite{Maze2008,Balasubramanian2008,Lange2011}. These results make the NV defect an appealing candidate for interfacing a nanomechanical oscillator: once immersed in a strong magnetic field gradient, an efficient coupling between the NV defect electronic spin and the nanoresonator position can be achieved. Furthermore, this novel hybrid system is susceptible to reach the strong coupling regime, as already envisioned in ref. \cite{Rabl2009,Rabl2010}.\\
In the following, we first show that the nanomechanical oscillator dynamics can be probed using the NV center as a single photon source, illustrating a resonant optomechanical coupling that does not suffer from the usual reduction in strength typically observed while optically interacting with sub-wavelength sized resonators. Furthermore we provide clear spectroscopic evidence of the mechanical degree of freedom by magnetic coupling of the spin to the nanoresonator position, demonstrating spin mediated readout of the oscillator dynamics.\\

The nanomechanical oscillator consists of a SiC nanowire attached to the extremity of a conducting tungsten tip (fig. 1). SiC nanowires represent compelling nanomechanical oscillators, combining in a low mass system a high mechanical quality factor, a relatively high vibration frequency and a large spreading of the zero-point energy wave function. For a $10\,\rm \mu m$ long and $50\,\rm nm$ diameter nanowire, with an effective mass of $M_{\rm eff}=16\,\rm fg$, the vibration frequency reaches $\Omega_{\rm m}/2\pi\equiv 1/T=1\,\rm MHz$ and its spring constant $k=1/M_{\rm eff}\Omega_{\rm m}^2=700\,\rm\mu N/m$, corresponding to a room temperature Brownian motion of 3\,nm rms and a groundstate wave function spreading of $\Delta x^q=\sqrt{\frac{\hbar}{2 M_{\rm eff}\Omega_{\rm m}}}\approx 0.7\,\rm pm$. The  nanomechanical oscillator can be efficiently driven to large oscillation amplitudes  (several $\mu m$),  as shown in fig. 2a. Resonators with a mechanical quality factor (Q) above 10 000 were measured in vacuum in the TEM imager, and even larger values can be achieved in similar devices \cite{Perisanu2007}.\\

A diamond nanocrystal hosting a single NV defect is attached to the oscillator free extremity and fluorescence is detected by a confocal microscope as shown in fig.~1 (see Methods).
When the hybrid system is set into motion, its extremity oscillates back and forth across the optical spot, thus modulating the overlap with the optical detection volume (fig.~2b). As the emitter can only be pumped and a photon detected when located within this volume, the fluorescence rate therefore provides a simple detection technique of the resonator dynamics (fig 2c). A time-resolved fluorescence measurement synchronized with the piezo driving voltage (fig.~2d) probes the oscillator dynamics across the optical spot. From this, the oscillation amplitude and direction can be determined and the piezo driving efficiency calibrated (of the order of 200 nm/V for the 625 kHz mode considered here).
Note that in this experiment, the NV center serves as a probe of the nanoresonator dynamics, and the recoil displacements $\delta x^{\rm rec}=\frac{h/\lambda}{M_{\rm eff}\Omega_{\rm m}}\approx 10\,\rm am$  due to single photon emission ($\lambda=638\,\rm nm$)  are negligible compared to $\Delta x^q$, which is equivalent to the Lamb-Dicke regime \cite{Blatt2008}.  \\

\begin{figure}[t]
\begin{center}
\includegraphics[width= \linewidth]{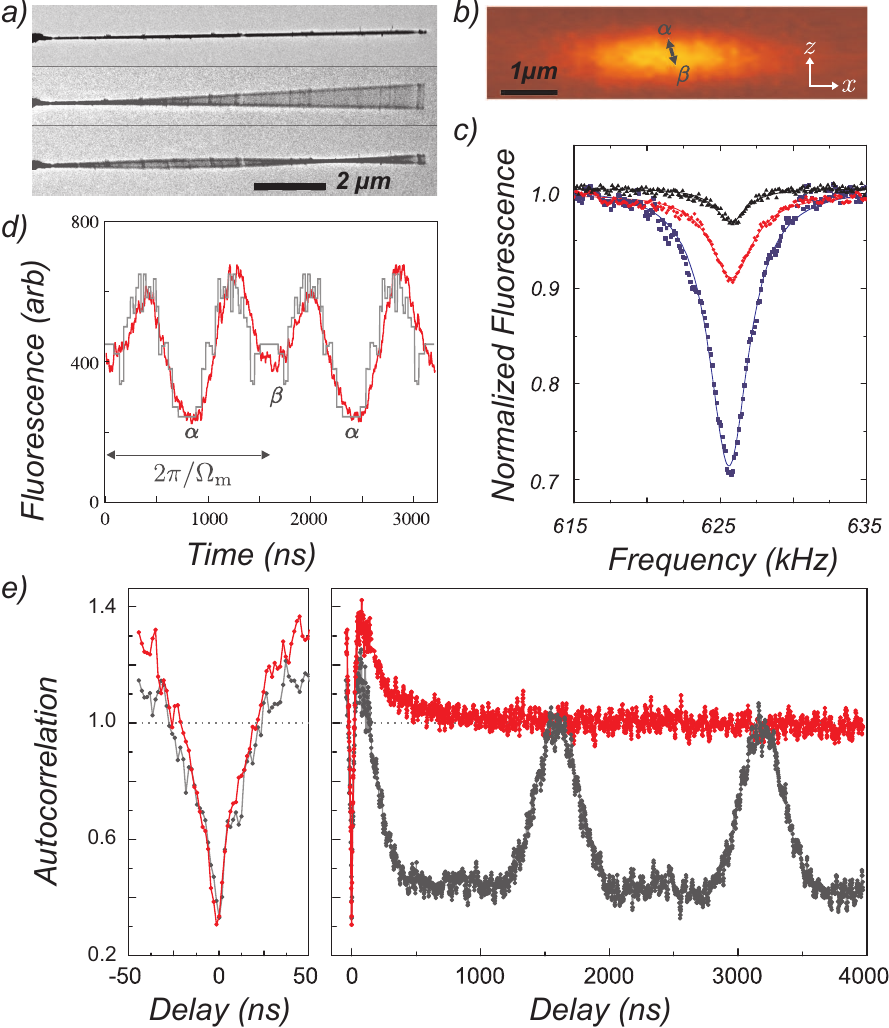}
\caption{
{\small A single quantum emitter with a mechanical degree of freedom  (a): Transmission electron microscope images of a SiC nanowire at rest (top) and  oscillating under electrostatic excitation at the frequency of its 2 lower eigenmodes (654 and 4887 kHz). (b):  Fluorescence map of the suspended NV within the optical spot in its oscillation plane. (c): Detection of piezo driven nanomechanical resonances by monitoring the mean NV fluorescence as a function of the piezo driving frequency for increasing amplitudes (from top to bottom: 50, 100 and 200\,nm at resonance). (d): In red, time-resolved fluorescence of the suspended NV center driven at 625\,kHz. Its comparison to the static fluorescence map of panel (b) allows one to extract the oscillation  parameters (i.e. direction, amplitude and turning points ($\alpha, \beta$)), represented in grey in panels (b, d). (e): Photon correlation measurements, recorded using a standard HBT setup, of the suspended single NV center at rest (red) and oscillating with a $200\,\rm nm$ amplitude at 625 kHz (black).}\\}
\end{center}
\end{figure}

The mechanical degree of freedom given to the single quantum emitter is further elucidated by recording the histogram of the time delays between two consecutive single-photon detections using a standard Hanbury Brown and Twiss (HBT) interferometer. After normalization to a Poissonnian statistics \cite{Beveratos2002}, the recorded histogram is equivalent to a measurement of the second-order autocorrelation function $g^{(2)}(\tau)$. For an oscillator at rest, a pronounced anticorrelation  effect  is observed ($g^{(2)}(0)=0.3$), as expected for a single quantum emitter (fig.~2e). The shape of the autocorrelation function is strongly altered when the emitter is in motion. Although the anticorrelation effect is still observed at zero delay, additional periodic drops appear, reflecting the time intervals spent outside the detection volume, as usually observed for the $g^{(2)}(\tau)$ function recorded under pulsed excitation.
The regime presented here corresponds to a slow oscillator, driven at amplitudes for which the time required to cross the optical detection volume - corresponding to the width of the peaks in the autocorrelation trace - remains long compared to the single emitter lifetime (12 ns). The photon emission probability of the single emitter therefore adiabatically follows the spatial variations of the pump intensity as it traverses the optical spot. A different and equally interesting regime arises in the situation where the illumination duration is comparable to the emitter lifetime. This can be readily reached, e.g. with  a 10-MHz oscillator, driven at $1\,\rm \mu m$ oscillation amplitudes.\\
Together, the above presented results call upon a wide range of experiments merging the fields of single emitter quantum optics and optomechanics.

The second part of this letter demonstrates the coupling between the nanomechanical oscillator position and the NV defect electronic spin. The ground state is a spin triplet $S=1$, whose degeneracy is lifted to $2.8$ GHz by spin-spin interactions in the absence of static magnetic fields (fig. 1b) \cite{Manson2006}. Radiative transition selection rules associated with the spin state quantum number provide a high degree of spin polarization in the $m_{s}=0$ substate through optical pumping. In addition, the NV defect photoluminescence intensity is significantly higher when the $m_{S}=0$  state is populated \cite{Manson2006}. Due to this spin dependent fluorescence rate, electron spin resonances (ESR) can be optically detected \cite{Gruber1997,Jelezko2004}. More precisely, as shown in fig. 3a, when the suspended single NV defect, initially prepared in the $m_{S}=0$  state through optical pumping, is driven to the $m_{S}=\pm 1$ spin states by applying a resonant microwave field, a dip in the photoluminescence signal is observed. The orientation of the suspended NV defect was determined by measuring the Zeeman shift of the ESR frequencies as a function of the orientation and magnitude of a calibrated static magnetic field (fig. 3b). The latter were subsequently fitted according to the eigenvalues of the ground-state spin Hamiltonian given by $H_{\rm spin}= D S_Z^2+ E(S_X^2-S_Y^2)+ g\mu_B {\bf B}\cdot {\bf S}$,
where $D$ and $E$ are the zero-field splitting parameters,  $Z$ the NV defect quantization axis,  $g$ its g-factor $(\approx 2$), and $\mu_{B}$ the Bohr magneton. The NV axis was found to be aligned (within 5 degrees) with the oscillation trajectory of a $625\,\rm kHz$ mode of the nanoresonator, coinciding  with the $z$ axis of fig. 1a.\\

\begin{figure}[t]
\begin{center}
\includegraphics[width=.95 \linewidth]{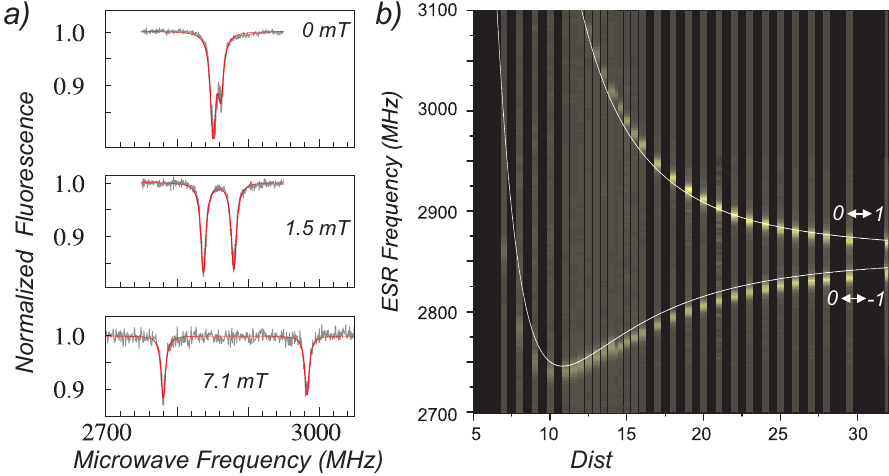}
\caption{ {\small  Optically detected ESR measured on the suspended NV center at rest. (a):  Zeeman shift of the ESR frequencies for increasing values of the magnetic field applied along an arbitrary direction. (b): Typical map of the ESR resonances as a function of the distance to a calibrated  magnet. The fit (white lines) gives the orientation of the NV axis.}\\}
\end{center}
\end{figure}

To magnetically couple the electronic spin and the nanoresonator position we apply a strong magnetic field gradient to the suspended NV, rendering its electronic spin energy dependent on the oscillator position $z$.
To this end an in-house patterned magnetic structure \cite{Kustov2010}  with an extended homogeneity of the field gradient was micro-positioned in the vicinity of the suspended NV. The magnetic field was aligned with the NV axis in order to maintain a high ESR contrast and the position optimized to find a gradient being homogeneous along the oscillating NV trajectory.
A prominent signature of the coupling is the modification of the ESR profile when the oscillator is set in motion. Since the oscillation frequency of the mode considered here ($ 625 \,\rm  kHz$) is smaller than the ESR linewidth (power broadened to a half-width at half maximum (HWHM) of $\Gamma_{\rm s}/2\pi = 7\,\rm MHz$), we can consider that the electron spin adiabatically follows the Zeeman shifted resonances. The evidence of magnetic coupling between the nanomechanical oscillator position and the NV electronic spin is illustrated in fig. 4b, where one can observe a motional ESR broadening followed by a characteristic splitting at stronger oscillation amplitudes ($\delta z$), whose shape reflects the harmonic oscillation turning points.

For a NV axis oriented along the oscillation direction and magnetic field ($z\simeq Z$), which holds true in our system, the  system is formally described by the coupling Hamiltonian $g\mu_{\rm B}\nabla B\, S_Z \,  z $. In this case, we can  approximate the magnetic coupling by a scalar description. Accordingly, the data from fig.\,4b were fitted with the function
$$\Lambda(f,\delta z)= \frac{1}{T}\int_0^{T}{\mathfrak{L}\left(f,f_0-\frac{g\mu_{\rm B}}{h} B(\delta z \, \cos \Omega_{\rm m}t)\right)dt},$$
where $\mathfrak{L}(f,f_{\rm ESR})$ is the unperturbed ESR resonance shape, which for simplicity is supposed to be  Lorentzian  (half-width $\Gamma_{\rm s}/2\pi$).  $B(z)$ is the magnetic field  along the NV trajectory, approximated by  $B(z)=B_0 + \frac{d B}{dz} z + \frac{d^2 B}{dz^2} z^2 $, whose coefficients are the only free parameters in the fitting procedure.
This model is in good agreement with experimental data, even at strong oscillation amplitudes. The effective spin resonance half-width defined as $\frac{1}{2\pi}\sqrt{\Gamma_{\rm s}^2+\left(\frac{g\mu_{\rm B}}{\hbar}\frac{d B}{dz}\delta z\right)^2}$ is then plotted as a function of the calibrated oscillation amplitude $\delta z$ in fig. 4c (in red for  data set presented in 4b, in blue and black for different gradients). The plot allows one to verify the consistency of each data set and to extract the magnetic field gradient which amounts to 6700 T/m for the data shown in fig. 4b. This value, as well as the mean magnetic field also obtained from the fit ($B_0 = 90\,\rm mT$) are in good agreement with both static measurements of the field spatial profile obtained by locally displacing the magnetic structure, as well as simulations \cite{Kustov2010}. The mechanical quality factor  (damped to $Q=150$ in air) is obtained by sweeping the driving frequency across the resonance (fig. 4d).

\begin{figure}[t]
\begin{center}
\includegraphics[width=\linewidth]{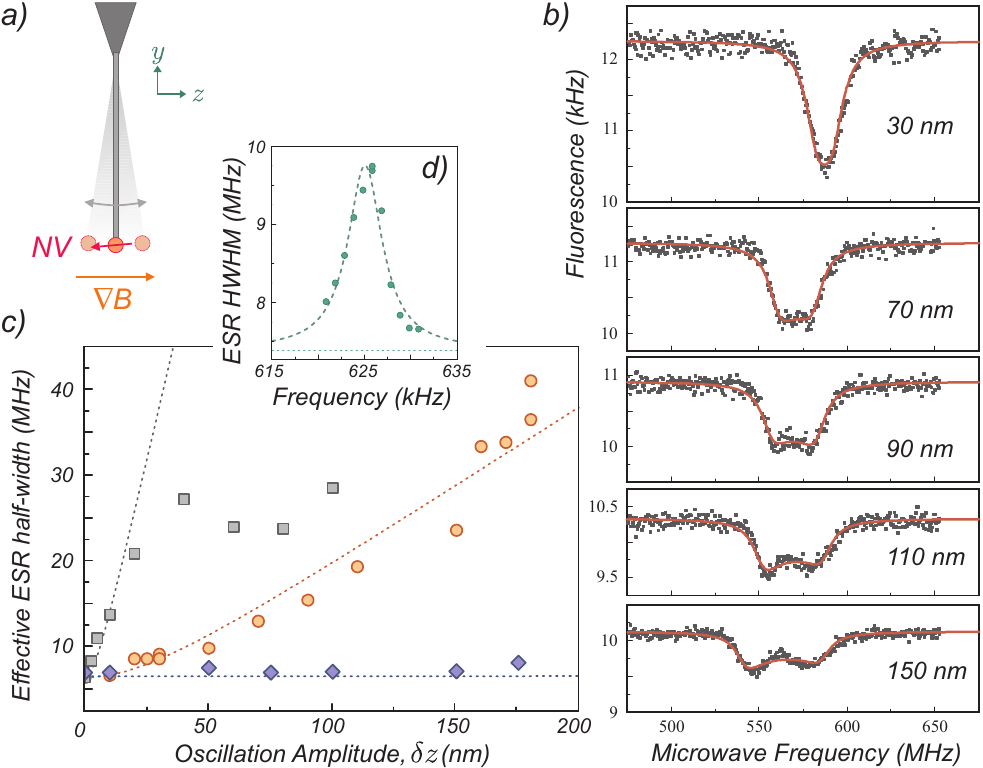}
\caption{{\small Magnetic coupling of the NV electronic spin to the nanomotion observed on the $m_{\rm S}=0$ to $m_{\rm S}=-1$ transition. (a): Schematics of the experiment. (b): ESR obtained for an increasing oscillation amplitude at $\Omega_{\rm m}/2\pi= 625\,\rm kHz $. The red line is a fit (see main text) allowing one to extract the effective ESR half-width which is reported in panel (c)  (circles) as a function of the oscillation amplitude ($\delta z$). (c): Various magnetic field gradients (0.1, 6700 and 45\,000 T/m) were explored, corresponding to different distances above the magnetic structure (1000, 15 and $2\,\rm\mu m$ (blue, red, black)), allowing  to tune the NV-oscillator coupling strength. The deviation from the model observed at large driving amplitudes in the strongest gradient case (black squares) is a consequence of the
reduced gradient homogeneity at short distances from the structure. (d):  Effective ESR half-width as a function of the driving frequency, for a 4500\,T/m gradient and a resonant oscillation amplitude of 50\,nm.
}}
\end{center}
\end{figure}

Having observed how the nanomotion is imprinted on the electronic spin dynamics, a question that naturally arises is whether the NV electronic spin can affect the nanoresonator dynamics \cite{Rugar2004}. This reverse interaction would enable cooling of the nanoresonator or preparation of non-classical mechanical states through spin dependent forces \cite{Blatt2008}. For a magnetic field gradient of $10^5\,\rm T/m$, the change in the spin dependent force exerted on the nanomechanical oscillator from one spin state to another amounts to $g \mu_B \nabla B \approx 2\,\rm aN$. This order of magnitude is comparable to the thermal noise limited force sensitivity of the nanoresonator $\sqrt{2 M_{\rm eff} \Omega_{\rm m} k_B T /{Q}}\approx  9\,\rm aN/\sqrt{Hz}$ expected at room temperature for the parameters previously used and the vacuum $Q=10\,000$.
Resolving the Brownian motion thus gives a reasonable metric for the sensitivity required to detect the spin dynamics. The corresponding room temperature thermal noise amounts at resonance (1 MHz) to approx. $100\,\rm  pm/\sqrt{Hz}$, an order of magnitude that can be easily detected with simple optical means despite the sub-wavelength size of the resonator \cite{Sanii2010,Favero2009,Anetsberger2009,Regal2008}.
Furthermore, to probe spin dynamics with the nanoresonator, spin coherence has to be preserved over several mechanical oscillations. The so-called resolved sideband regime ($\Omega_{\rm m}>\Gamma_{\rm s}$) is within reach when working with shorter nanowires and increased spin coherence times, and is also of importance when exploring the avenues for probing and cooling the nanomechanical oscillator down to its quantum ground state through single spin manipulations \cite{Rabl2009,Rabl2010a}.

These results represent a clear quantitative signature of the nanoresonator motion directly imprinted on the electronic spin dynamics via magnetic coupling.
Long lived electronic spins  coupled to nanomechanical oscillators represent a promising experimental hybrid system whose two components can independently be monitored and controlled. This, combined with the single photon source character of the suspended NV defect paves the way towards single photon optomechanics.\\

{\bf Acknowledgments }
We acknowledge  J. Jarreau, C. Hoarau, D. Lepoitevin, J.F. Motte, P. Brichon, N. Dempsey, O. Fruchart, F. Dumas Bouchiat, D. Givord, E. Gheeraert, O. Mollet, A. Drezet, J.F. Roch, S. Huant and J. Chevrier for technical support, experimental assistance and discussions. This project is funded by the European Commission (Reintegration Grant) and the Agence Nationale de la Recherche (project Q-NOM).\\

{\bf Methods }

{\it Nanowire functionalization -} A diamond nanocrystal hosting a single NV defect is attached to the oscillator free extremity during a piezo-controlled immersion into a commercial solution of 50-nm-diameter diamond nano-crystals. The adhesion efficiency is significantly increased under focussed laser illumination, due to the increased convection combined with an optical tweezer mechanism. Since the solution meniscus size remains comparable to the nanowire diameter, it is possible to only functionalize the very extremity of the nanowire, while a  subsequent focussed ion beam cut allows final adjustments. This method allows for efficient and robust positioning of a single NV center at the extremity of a nanowire  and works reliably over a wide variety of resonator sizes and materials including Carbon and Boron Nitride nanotubes.

{\it Experimental setup - }
The NV center is excited and its fluorescence collected through a 100x long working distance microscope objective (generating an approx. 450\,nm diameter optical spot) and detected on avalanche photodiodes. The objective is mounted on a fast Physik Instrument XYZ piezo stage in order to localize the suspended NV defect (fig. 1c). A tracking program continuously maintains the detection spot on the single emitter.
A fast piezoelectric module positioned on top of the STM tip drives the nanomechanical oscillator and a micro-antenna generates the microwave field used to manipulate the NV electronic spin.

%\bibliography{biblio-NV}

\end{document}